\documentclass[aip,jcp,reprint]{revtex4-1}

\usepackage{chemformula} 
\usepackage[T1]{fontenc} 
\usepackage{amsmath}
\usepackage{graphicx}
\usepackage{bm}
\usepackage{mathrsfs}
\usepackage{subfigure}
\usepackage{xcolor}
\usepackage{float} 
\usepackage{xr}
\usepackage{tikz}
\usepackage{soul} 
\usepackage{color, xcolor} 
\usepackage{booktabs}
\usepackage{chemarrow}
\usepackage{rviewport}
\usepackage{dcolumn}
\usepackage{multirow}
\usepackage{threeparttable}

\newcommand*\op[1]{\hat{#1}}
\newcommand{\Hs}{\hat{H}_{\rm s}}
\newcommand{\Hph}{\hat{H}_{\rm ph}}
\newcommand{\Hsph}{\hat{H}_{\rm s-ph}}
\newcommand{\Ueff}{U_{\mathrm{eff}}}
\newcommand{\DyAC}{\textbf{DyAC}}
\newcommand{\DyCp}{\textbf{DyCp}}
\newcommand{\DybBr}{\textbf{DybBr}}

\newcommand{\set}[1]{\{#1\}}

\newcommand{\twocol}[1]{\multicolumn{2}{c}{#1}}

\newcommand{\SI}[1]{{\color{blue} #1}}

\begin{document}

\title{Accelerating \emph{ab initio} spin-phonon relaxation simulation of single-ion magnets by quantum embedding and spatial truncation}

\author{Yifan Deng}
\author{Zhe-Bin Guan}
\author{Zilong Zou}
\author{Zheng Sun}
\author{Ze-Wei Li}
\author{Bingwu Wang}
\email{wangbw@pku.edu.cn}
\author{Hong Jiang}
\email{jianghchem@pku.edu.cn}
\affiliation{Beijing National Laboratory for Molecular Sciences, College of Chemistry and Molecular Engineering, Peking University, Beijing, 100871}

\date{\today}

\begin{abstract}
Single-ion magnets (SIMs) show promise for high-density storage and quantum computing, but predicting spin-phonon coupling (SPC) and magnetic relaxation remains challenging due to the need for numerous non-equilibrium multiconfigurational calculations. Recent advances in quantum embedding methods offer a potential route to address this issue. In this work, density matrix embedding theory (DMET) combined with complete active space self-consistent field (CASSCF) is benchmarked for the static magnetic properties and spin-phonon coupling (SPC) parameters of Dy$^{3+}$-based SIMs. The method is further combined with spatial truncation to calculate SPC parameters for these SIMs. It is found that truncating the space near the first coordination sphere reduces the computational cost dramatically while keeping the errors in the effective energy barrier and relaxation time-scale negligible. This study provides a practical calculation framework for accurate and efficient spin dynamics prediction, laying the foundation for the rational design of high-performance single-molecule magnets.
\end{abstract}

\maketitle

\section{Introduction \label{sec:intro}}

Single-molecule magnets (SMMs) are a class of molecular compounds that exhibit slow magnetic relaxation and magnetic hysteresis due to their intrinsic magnetic anisotropy.\cite{Christou2000, Gatteschi2006} Their defining features, i.e. a high effective energy barrier (\(U_{\mathrm{eff}}\)) for magnetic moment reversal and a long relaxation time (\(\tau\)), render them highly attractive for applications in ultra-high-density information storage,\cite{Maehrlein2018} spintronic devices,\cite{Vzutic2004, Bogani2008} quantum information processing,\cite{Wolfowicz2021, Atzori2019} and beyond.\cite{Coronado2020, Chilton2022} Among the diverse SMMs developed to date,\cite{Feng2018, FerrandoSoria2017} lanthanide-based single-ion magnets (Ln-SIMs) have garnered considerable attention in recent years, owing to the remarkable progress in enhancing their \(U_{\mathrm{eff}}\) and blocking temperatures (\(T_{\mathrm{B}}\)).\cite{Ishikawa2003, JiangSD2010, JiangSD2011, Rinehart2011, Gupta2016, Goodwin2017, Demir2017, Guo2018, Velkos2019, Emerson-King2025} However, the practical applicability of Ln-SIMs is critically constrained by the limited magnetic relaxation time, for which spin-phonon coupling plays a dominant role. Indeed, the relaxation dynamics in these systems are governed largely by the interaction between an effective or pseudo-spin degree of freedom and structural vibration (phonon), which facilitates energy dissipation and limits the lifetime of the magnetized state.\cite{Escalera-Moreno2018, Lunghi2020AMR, Chilton2022} As a result, a thorough understanding of spin-phonon relaxation processes has become a central theme in SMM research, not only for rationalizing observed magnetic behavior but also for guiding the design of systems with enhanced performance. This focus is particularly pertinent given that spin-phonon coupling is often the primary factor limiting both \(U_{\mathrm{eff}}\) and \(T_{\mathrm{B}}\) under practical conditions, thus making it a vital target for both experimental and theoretical investigation.

Quantum chemical calculations rooted in multi-configurational wave function theory (MC-WFT), particularly the complete active space self-consistent field (CASSCF) approach to scalar-relativistic Hamiltonians followed by state interaction treatment of spin-orbit coupling (SOC), abbreviated as CASSCF-SO henceforth, have been instrumental in advancing fundamental research on transition metal and lanthanide SIMs.\cite{Atanasov2015} These methods afford unprecedented microscopic insights into the electronic structure, magnetic anisotropy, and spin dynamics of transition-metal (3d) and lanthanide (4f) systems.\cite{Ungur2015, Lunghi2022NRC, Chibotaru2023} A major methodological breakthrough emerged about a decade ago with the development of \textit{ab initio} spin-dynamics approaches, which combine the \textit{ab initio} calculation based effective Hamiltonian with open quantum systems theory.\cite{Lunghi2017NC, Escalera-Moreno2017, Goodwin2017} This framework has since been widely adopted to compute spin-relaxation rates and disentangle the underlying relaxation mechanisms in a variety of SIMs.\cite{Lunghi2017CS, Lunghi2017NC, Escalera-Moreno2018, Evans2019, YuKX2020, Reta2021, Briganti2021, Lunghi2022SA, Mondal2022}

However, the practical implementation of such approaches comes at a formidable computational price. Extracting reliable spin-phonon coupling (SPC) parameters requires performing \textit{ab initio} calculations across a large ensemble of off-equilibrium molecular geometries to sample the vibrational landscape adequately, a task that rapidly becomes prohibitive, especially for systems with many degrees of freedom. This computational bottleneck not only limits the throughput of relaxation-rate predictions but also hampers rapid screening and rational design of new SIM candidates. In response, a variety of strategies have been put forward to alleviate these costs. Notable examples include semi-\textit{ab initio} treatments that incorporate acoustic phonon contributions,\cite{Garlatti2021} linear vibronic coupling models combined with analytical CASSCF gradients for efficient SPC evaluation,\cite{Staab2022, Nabi2023} and, more recently, machine-learning techniques that enable swift and accurate prediction of magnetic anisotropy tensors and SPC parameters directly from molecular structures.\cite{Lunghi2022SA, Lunghi2020JCP, Lunghi2020JPCC, Lunghi2020JPCL, Zaverkin2022, Nguyen2022, Briganti2025} These evolving methodologies collectively signal a shift toward more accessible and predictive computational frameworks, while still preserving the essential quantum-chemical rigor required for reliable spin-dynamics simulations.

The past decade has witnessed the emergence of various quantum embedding approaches\cite{SunQ2016, Jones2020, Vorwerk2022} aimed at reducing the computational cost of high-level quantum chemistry methods like CASSCF\cite{Roos1980} for complex systems. These methods share a common strategy: partitioning the system into a fragment of interest (the ``impurity'') and its environment.\cite{SunQ2016, Jones2020} A low-level method like Hartree-Fock (HF) or Kohn-Sham density-functional theory (KS-DFT) is applied to the full system to construct an effective Hamiltonian for the \textit{embedded impurity}, which is then treated by a high-level method. Among the various embedding formalisms developed based on electron density,\cite{Jacob2014, Wesolowski2015} density matrix,\cite{Knizia2012, Fornace2015, Welborn2016, Yu2017} or Green's functions\cite{Kotliar2006, Zgid2017, Ma2021}, density matrix embedding theory (DMET), proposed by Chan and coworkers \cite{Knizia2012, Knizia2013, Wouters2016}, stands out due to its rigorous formulation and the provision of an embedded impurity Hamiltonian in the second-quantization representation, enabling seamless integration with various high-level solvers. Over the past decade, DMET and its variants \cite{Welborn2016, Fertitta2019, Nusspickel2022, Nusspickel2023, Sekaran2023, Yalouz2022} have attracted considerable interest and demonstrated excellent performance across a range of chemical systems \cite{Wouters2016, Bulik2014, Pham2018, Mitra2022, CuiZH2020b, CuiZH2022, Nusspickel2022, Haldar2023, AiY2022, CaoC2023, GuanZB2025, AiY2025, SunX2025, SunY2025, Verma2026}.

The effective spin Hamiltonian, which plays a central role in the theoretical modeling of SIMs \cite{Chibotaru2013, Atanasov2015, Neese2002}, is predominantly governed by the interactions between the central metal ion and its coordinating ligand environment. This inherent electronic structure naturally lends itself to an embedding-based theoretical description, wherein the metal center is treated as the impurity and the ligands as the surrounding environment. Following this rationale, we have recently developed a DMET-based quantum embedding approach specifically tailored for SIMs,\cite{AiY2022, AiY2025} employing restricted open-shell Hartree-Fock (ROHF) as the low-level solver and CASSCF-SO as the high-level solver, hence termed DMET+CASSCF-SO. As demonstrated in our previous works,\cite{AiY2022, AiY2025, GuanZB2025} this method accurately reproduces the energy level splittings and zero-field splitting parameters of both 3d and 4f SIMs, while achieving a marked reduction in computational expense relative to the conventional all-electron CASSCF-SO treatment.

In this work, we propose to combine DMET+CASSCF-SO and a spatial truncation scheme to accelerate \textit{ab initio} calculation of SPC parameters, which can dramatically reduce the computational cost of ab initio spin-phonon relaxation dynamics simulation with little loss of accuracy. The paper is organized as follows. Sec. \ref{sec:method} gives a brief introduction to theoretical methods used in this work and some computational details. Sec. \ref{sec:results} presents our main results, including a validation of the accuracy of DMET+CASSCF-SO in the description of effective spin Hamiltonian parameters and spin-phonon coupling parameters, and a detailed comparison for the spin relaxation time calculated with DMET and spatial truncation and that from the full treatment. Sec. \ref{sec:conclusion} summarizes our main findings and remarks on possible extensions of this work in the future.

It should be noted that when we were preparing this manuscript, Verma et al. presented a related work that applied multi-reference DMET approach to spin-phonon relaxation of 3d and 4f SIMs \cite{Verma2026arXiv}.

\section{Theoretical Methods and Computational Details \label{sec:method}}

\subsection{Density-matrix embedding theory}

We give a brief introduction to the essential features of density matrix embedding theory (DMET), and more detailed formulations can be found in Refs. \citenum{Wouters2016, Wouters2017, Verma2026}. DMET, like many other quantum embedding approaches \cite{SunQ2016, Jones2020}, starts with partitioning the full system into the fragment of interest, also termed as the impurity, and its surrounding environment, denoted as $\mathcal{I}$ and $\mathcal{E}$, respectively, in terms of atom-centered localized orthogonal orbitals (LOs). For the Ln-SIMs considered in this work, the impurity is chosen as all L{\"o}wdin-orthogonalized atomic orbitals centered on the central lanthanide ion. The many-body interaction between the impurity and the environment is described by a set of so-called bath orbitals, which are linear combinations of environment LOs and constructed based on Schmidt decomposition of the ground state Hartree-Fock Slater determinant \cite{Knizia2012}. These bath orbitals are combined with the impurity orbitals to span an embedded impurity space, denoted as $\mathcal{I}_{\rm emb}$, which is usually much smaller than the full space of the system. The remaining environment orbitals are regarded as ``unentangled'' with the impurity \cite{Wouters2016}, and can be further partitioned into occupied (often termed as core orbitals) and unoccupied ones, denoted as $\mathcal{U}_{\text{occ}}$ and $\mathcal{U}_{\text{vir}}$, respectively. Projecting the Hamiltonian of the full system to $\mathcal{I}_{\rm emb}$, one obtains the following embedded impurity Hamiltonian in the second quantization form,
\begin{equation}
\begin{aligned}
    \op{H}_{\mathrm{emb}} &= \sum_{i,j \in \mathcal{I}_{\rm emb}} \sum_{\sigma} \tilde{h}_{ij} \op{c}_{i\sigma}^\dagger \op{c}_{j\sigma} \\
    &+ \frac{1}{2} \sum_{i,j,k,l \in \mathcal{I}_{\rm emb}} \sum_{\sigma\sigma'} \langle ij | kl\rangle \op{c}_{i\sigma}^\dagger \op{c}_{j\sigma'}^\dagger \op{c}_{l\sigma'} \op{c}_{k\sigma}.
\end{aligned}
\end{equation}
with $\langle ij | kl\rangle \equiv \int \int d\mathbf{r_1}d\mathbf{r_2} \phi_i^*(\mathbf r_1) \phi_j^*(\mathbf r_2) r_{12}^{-1} \phi_k(\mathbf r_1) \phi_l (\mathbf r_2) $ and $\tilde{h}_{ij}$ being the matrix element of the following effective single-particle operator
\begin{equation}
    \op{\tilde{h}} = -\frac{1}{2} \nabla^2 - \sum_{I} \frac{Z_I}{|\mathbf{r} - \mathbf{R}_I|} + \sum_{a\in \mathcal{U}_{\mathrm{occ}}} (2\op{J}_a - \op{K}_a)
\end{equation}
where $\op{J}_a$ and $\op{K}_a$ are the Coulomb and exchange operators of core orbital $a$, respectively, and the summation in the second term is over all nuclei in the system. With its greatly reduced degrees of freedom, the embedded Hamiltonian can be solved by high-level quantum chemistry solvers like CASSCF-SO with greatly reduced computational cost \cite{AiY2022, AiY2025, GuanZB2025}.


\subsection{Ab initio theory for spin-phonon relaxation}

The spin-phonon relaxation dynamics of SMMs is commonly described within the framework of open quantum systems,\cite{Breuer2002, Lunghi2022SA, Kragskow2023} where the total Hamiltonian that accounts for both spin and phonon degrees of freedom is partitioned into three parts
\begin{equation}\label{eq:Htot}
	\hat{H} = \Hs+\Hph+\Hsph,
\end{equation}
corresponding to the spin and phonon subsystems and their mutual interaction, respectively.\cite{Breuer2002, Lunghi2022SA} For lanthanide-based SIMs, the spin Hamiltonian is typically expressed in terms of an effective crystal-field Hamiltonian acting within the ground state SOC multiplet \cite{Chibotaru2013, Chibotaru2023, Chilton2025}
\begin{equation}
	\Hs = \sum_{k=2,4,6} \sum_{q=-k}^k B_{kq} \hat{O}_{kq} (\hat{\mathbf{J}}), \label{Con:spinh}
\end{equation}
where \(\hat{O}_{kq}(\hat{\mathbf{J}})\) are tesseral tensor operators corresponding to the total angular momentum operator \(\hat{\mathbf{J}}\), and \(B_{kq}\) are the effective crystal-field parameters (CFPs) that encapsulate the splitting of the ground state multiplet in the coordinating ligand field.\cite{Ryabov2009, Rudowicz2015} The phonon Hamiltonian \(\Hph\) describes the quantized vibrational modes of the molecular crystal, which within the harmonic approximation take the form of independent oscillators with frequencies \(\set{\omega_\nu}\) and occupation numbers governed by the Bose-Einstein distribution.\cite{Kragskow2023}

The spin-phonon coupling arises from the modulation of the effective crystal-field Hamiltonian by nuclear vibration. To first order in the normal-mode coordinates \(\set{Q_\nu}\), the interaction Hamiltonian can be written as a Taylor expansion of \(\Hs\) around the equilibrium geometry \cite{Goodwin2017, Kragskow2023}
\begin{equation}
	\Hsph = \sum_{\nu}\left(\frac{\partial\Hs}{\partial Q_\nu }\right) Q_\nu = \sum_{k,q,\nu} B_{kq,\nu}\,\hat{O}_{kq}\, Q_\nu, \label{Con:PH}
\end{equation}
where the coefficients \(B_{kq,\nu} \equiv \partial B_{kq}/\partial Q_\nu\) are the spin-phonon coupling (SPC) parameters that quantify the sensitivity of each CFP to a given vibrational distortion.\cite{Kragskow2023} In practice, these derivatives are often obtained by finite-difference calculations at displaced geometries based on CASSCF-SO.\cite{Staab2022, Kragskow2023}

Under the common assumption that the phonon bath equilibrates on a much faster timescale than the spin system, an assumption that has been validated for typical molecular crystals of lanthanide SMMs\cite{Kragskow2023}, the spin dynamics can be described by the Redfield equation or, upon secular approximation, by the Pauli master equation.\cite{Nitzan2006, Lunghi2023} In the latter framework, which we adopt in this work, the magnetic relaxation time is determined by the transition rates \(\gamma_{mn}\) between eigenstates \(|m\rangle\) and \(|n\rangle\) of the crystal-field Hamiltonian. Restricting to single-phonon (Orbach) processes, these rates take the form \cite{Goodwin2017, Briganti2021, Lunghi2023, Kragskow2023}
\begin{equation} \label{eq:Orbach}
	\gamma_{mn}^{\rm 1-ph}=\frac{\pi}{\hbar^2}\sum_{\nu}|V^{\nu}_{mn}|^2\,G^{\rm 1-ph}(\omega_{mn},\omega_{\nu}),
\end{equation}
where \(V^{\nu}_{mn} \equiv \langle m|\partial\Hs/\partial Q_\nu |n\rangle\) is the matrix element of the first-order spin-phonon coupling operator between spin states, and the kernel $G^{\rm 1-ph}$ takes the form
\begin{equation} \label{eq:Green}
	G^{\rm 1-ph}(\omega_{mn},\omega_{\nu}) = \delta(\omega_{mn}-\omega_{\nu})\,\bar{n}_{\nu} + \delta(\omega_{mn}+\omega_{\nu})\,(\bar{n}_{\nu}+1),
\end{equation}
with \(\omega_{mn} \equiv (E_m - E_n)/\hbar\) denoting the energy splitting between spin states, and \(\bar{n}_{\nu} = [\exp(\hbar\omega_{\nu}/k_{\mathrm{B}}T) - 1]^{-1}\) representing the thermal occupation of phonon mode \(\nu\) according to Bose-Einstein statistics. The two terms in Eq. \ref{eq:Green} correspond respectively to phonon absorption (which promotes the spin from \(|m\rangle\) to the higher-energy \(|n\rangle\)) and phonon emission (which drives the reverse transition). In practice, the Dirac delta functions are replaced by normalized Gaussian functions with a finite width, commonly taken as 10 cm\(^{-1}\), to account for phonon lifetime broadening and to facilitate numerical integration over the dense phonon density of states.\cite{Lunghi2022SA, Kragskow2023}

With the theoretical framework established, the practical computation of transition rates hinges on the prior determination of the SPC parameters \(\set{B_{kq,\nu}}\). These parameters quantify how atomic displacements along a given normal mode perturb the crystal-field potential experienced by the metal ion. Within the harmonic approximation, the SPC parameters can be evaluated from the Cartesian derivatives of the CFPs via the following transformation \cite{Mondal2022, Kragskow2023}
\begin{equation}
	B_{kq,\nu} = \sum_{i}^{N_{\rm at}}\sum_{\alpha=x,y,z}\sqrt{\frac{\hbar}{2\omega_\nu m_i}}\,L_{i\alpha, \nu}\left(\frac{\partial B_{kq}}{\partial X_{i\alpha} }\right),  \label{Con:SPCCd}
\end{equation}
where \(N_{\rm at}\) is the number of atoms in the crystallographic unit cell, \(\omega_\nu\) is the frequency of phonon mode \(\nu\), \(m_i\) is the mass of the \(i\)-th atom, \(L_{i\alpha,\nu}\) are the mass-weighted eigenvector components of the Hessian matrix in the $\nu$-th mode, and \(X_{i\alpha}\) denotes the Cartesian coordinate of atom \(i\) along direction \(\alpha\). This expression makes explicit that the SPC parameters are determined by the collective contribution of all atoms in the unit cell, weighted by the inverse square root of their masses and the mode frequencies.

In practice, however, evaluating Eq. (\ref{Con:SPCCd}) for all atoms in a molecular crystal, particularly when large supercells are required to capture the phonon dispersion accurately, can become computationally prohibitive.\cite{Kragskow2023} Fortunately, for lanthanide-based SIMs, a significant simplification is justified by the localized nature of the 4f orbitals: the crystal-field potential at the metal center is predominantly dictated by the nearest-neighbor ligand atoms, while atoms farther away exert only a weak electrostatic influence that decays with distance. This spatial locality implies that the summation in Eq. (\ref{Con:SPCCd}) can be truncated to include only those atoms within a finite cutoff radius \(R_{\mathrm{c}}\) around the central metal ion, without appreciably compromising the accuracy of the resulting relaxation rates. The truncated SPC parameters are thus given by
\begin{equation}
	B_{kq,\nu}(R_{\mathrm{c}}) = \sum_{i} \theta(R_{\mathrm{c}} - R_i)\sum_{\alpha=x,y,z}\sqrt{\frac{\hbar}{2\omega_\nu m_i}}\,L_{\nu,i\alpha}\left(\frac{\partial B_{kq}}{\partial X_{i\alpha} }\right),  \label{Con:SPCCdr}
\end{equation}
where \(R_i\) denotes the distance from the \(i\)-th atom to the central metal ion (here Dy\(^{3+}\)), and \(\theta(R_{\mathrm{c}} - R_i)\) is the Heaviside step function that restricts the summation to atoms satisfying \(R_i \le R_{\mathrm{c}}\). It follows directly from Eq. (\ref{Con:SPCCd}) that the full SPC parameters are recovered in the limit of an infinitely large cutoff, i.e. \(B_{kq,\nu} = \lim_{R_{\mathrm{c}}\to\infty} B_{kq,\nu}(R_{\mathrm{c}})\). The convergence of the spin-phonon coupling and the resulting relaxation times with respect to \(R_{\mathrm{c}}\) can therefore be monitored systematically, allowing one to balance computational efficiency against numerical accuracy in practical calculations.

\begin{figure*}[ht]
	\centering
	\includegraphics[width=0.25\linewidth]{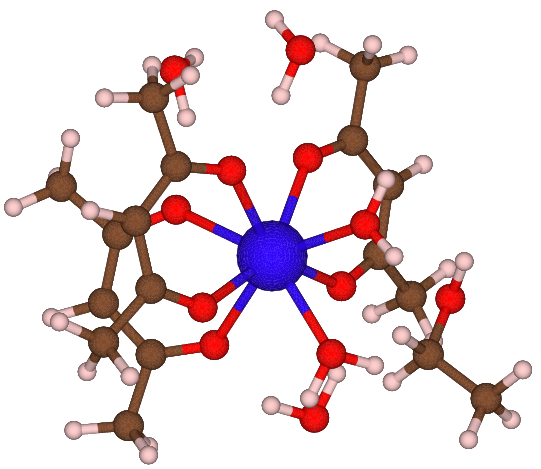}
	\includegraphics[width=0.25\linewidth]{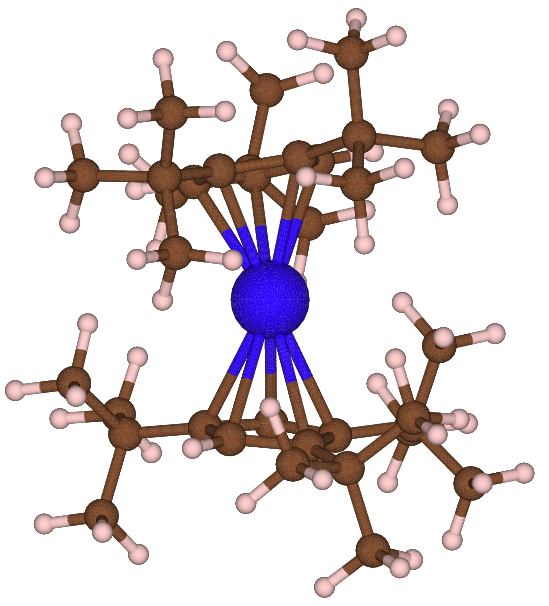}
	\includegraphics[width=0.25\linewidth]{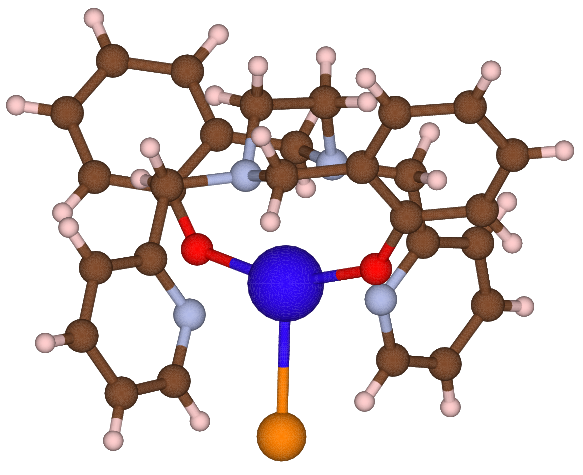}
	\caption{\label{fig:struct}
	Molecular structures of Dy$^{3+}$-based SIMs considered in this work.
	Left: \DyAC~($\mathrm {[Dy(acac)_3(H_2O)_2]}\cdot$ $\mathrm{EtOH\cdot H_2O}$, $\mathrm{acac}^-$ = acetylacetonate, EtOH = ethanol) \cite{JiangSD2010};
	Middle: \DyCp~($\mathrm{Dy[Cp^{ttt}_2][B(C_6F_5)_4]}$,$\mathrm{(Cp^{ttt} =C_5H_5Bu_3-1,2,4)}$)\cite{Goodwin2017};
	Right: \DybBr~($\mathbf{[Dy(bbpen)Br]}$, (bbpen= N, N’-bis(2-hydroxybenzyl)-N,N’-bis(2-methylpyridyl)ethylenediamine))\cite{Liu2016}. Color codes: Dy, dark blue; O, red; C, brown; Br, Orange; N, light blue; H, Salmon Pink.
	}
\end{figure*}

\subsection{Computational Details}

To benchmark the accuracy of our DMET+CASSCF-SO approach for spin-phonon relaxation, we selected three prototypical Dy\(^{3+}\)-based SIMs that feature distinct coordination environments, namely \DyAC \cite{JiangSD2010}, \DyCp \cite{Goodwin2017}, and \DybBr \cite{Liu2016}, as illustrated in Fig. \ref{fig:struct}. \DyAC\ features strong equatorial oxygen coordination, and exhibits field-induced slow magnetic relaxation with an effective barrier of $\Ueff = 130$ cm$^{-1}$ \cite{JiangSD2010}. \DyCp\ exemplifies the metallocene-based dysprosocenium systems that exhibit exceptional single-molecule magnet performances, showing an extremely large effective barrier of $\Ueff = 1223$ cm$^{-1}$ and a blocking temperature up to 60 K \cite{Goodwin2017}. \DybBr\ represents a  $D_{\rm{5h}}$ symmetry with axial oxygen and equatorial nitrogen halide coordination that displays a magnetization reversal barrier of $\Ueff = 712$ cm$^{-1}$ and the magnetic hysteresis up to 14 K \cite{Liu2016}. \SI{The geometric structures of these three systems used in the calculations of this work are given in the Supplementary Material}. For all three systems, the static effective Hamiltonians are computed using both the DMET+CASSCF-SO and the conventional all-electron CASSCF-SO methods. Subsequently, SPC parameters are evaluated using the DMET approach for all three complexes, while all-electron SPC calculations are performed only for \DyCp\ as a benchmark reference.

\textbf{Phonon calculations.} The phonon spectra of \DyAC\ and \DybBr\ were obtained from periodic density-functional theory (DFT) calculations using the Perdew-Burke-Ernzerhof (PBE) generalized gradient approximation (GGA),\cite{Perdew1996} as implemented in the VASP package.\cite{Kresse1996} Core-valence interactions were described by the projector augmented wave (PAW) method,\cite{Bloechl1994} with a plane-wave cutoff energy of 530 eV. Given the highly localized nature of the 4f electrons and their negligible influence on the structural and vibrational properties, an f-in-core PAW pseudopotential was employed for Dy\(^{3+}\) to circumvent the difficulties associated with the strong correlation of open-shell 4f states. For \DyCp, geometry optimization and vibrational analysis were carried out at the PBE-GGA level using the def2-SVP basis set for all atoms, as implemented in the ORCA package.\cite{Neese2020}

\textbf{Electronic structure and SPC calculations using DMET+CASSCF-SO.} Except for the all-electron SPC benchmark on \DyCp, all other \emph{ab initio} calculations of CFPs were performed using a locally extended version of the PySCF package.\cite{SunQ2018, SunQ2020} Our implementation of the DMET+CASSCF-SO method is available as an add-on module to PySCF, as is the module for computing CFPs, magnetic anisotropy \(\bm{g}\)-tensors, SPC parameters, and relaxation times. The DMET orbital selection threshold \(\epsilon\) was set to \(10^{-12}\) throughout\cite{AiY2022}. For state-averaged CASSCF (SA-CASSCF), an active space of (9e, 7o) was chosen for the Dy complexes, encompassing all 21 sextet states with equal weights in the state-averaging procedure. Scalar relativistic effects and spin-orbit coupling (SOC) were treated using the spin-free exact two-component theory in its one-electron variant (SFX2C-1e)\cite{Kutzelnigg2005, LiuWJ2009, Dyall2001} and the spin-orbit mean-field (SOMF) approximation to the Breit-Pauli Hamiltonian,\cite{Neese2005, Hess1996} respectively. The x2c-TZVPall-2c basis set\cite{Pollak2017} was employed for Dy, while x2c-SVPall-2c was used for all other atoms.

\textbf{All-electron SPC benchmark on \DyCp.} For the all-electron SPC reference calculations on \DyCp, the SPC parameters were derived from all-electron CASSCF-SO computations using the ORCA\cite{Neese2020} and Molforge\cite{Lunghi2022SA} programs. The SA-CASSCF settings were kept identical to those described above. Scalar relativistic effects were incorporated via the Douglas-Kroll (DK) transformation,\cite{Douglas1974, Hess1986} and SOC was treated using ORCA's default settings. The RIJCOSX approximation was employed for Coulomb and exchange integrals. The SARC2-DKH-QZVP basis set\cite{Aravena2016} was used for Dy, DKH-def2-TZVP\cite{Pantazis2008, Weigend2005} for C, and DKH-def2-SVP for H.

\textbf{SPC parameter evaluation.} In all SPC calculations, only \(\Gamma\)-point phonons were considered. The derivatives of the CFPs with respect to Cartesian displacements \(X_a\) were evaluated using a finite-difference scheme.\cite{Lunghi2017CS} Each molecular degree of freedom was sampled with 10 displacements evenly distributed over a range of \(\pm 0.1\) Å, with consecutive displaced structures along the same coordinate differing by a step size of 0.02 Å. This minimal step size enables the reuse of the ROHF wavefunction from the previous geometry as the initial guess for the next displacement, substantially reducing the overall computational overhead.

\section{Results and Discussion} \label{sec:results}
\begin{table}[ht]
	\centering
	\caption{\label{tab:Bkq} Effective crystal field parameters $B_{kq}$ (in cm$^{-1}$) corresponding to $k=2$ and the principal values of the effective $g$ tensors for three Ln-SIMs obtained from DMET and all-electron (AE) CASSCF-SO calculation. `gs' refers to the ground state Kramers doublets, and `mix' refers to the lowest strongly mixed Kramers doublets. The last three rows show the number of orbitals ($N_{\rm orb}$) used in DMET and all-electron CASSCF, the CPU time used in ROHF ($T_{\rm CPU}^{\rm ROHF}$) and subsequent CASSCF calculation ($T_{\rm CPU}^{\rm CAS}$), respectively.}
	\begin{threeparttable}
		\begin{tabular*}{0.48\textwidth}{@{\extracolsep{\fill}} ccccccc}
			\toprule
			           &\twocol{\DyAC}  &\twocol{\DyCp}   &\twocol{\DybBr}\\
			\cmidrule(lr){2-3} \cmidrule(lr){4-5} \cmidrule(lr){6-7}
			           & DMET   & AE    & DMET   & AE     & DMET  & AE \\
			\midrule
			$B_{2,-2}$ & -1.129 &-1.148 & 0.606  & 0.536  & 0.032 & 1.40e-4 \\
			$B_{2,-1}$ & 1.014  & 1.014 & -0.121 & -0.100 & 0.004 & 2.87e-5 \\
			$B_{2,0}$  & -3.355 & -3.324& -20.70 & -20.66 & -12.01 & -11.86 \\
			$B_{2,1}$  & -0.833 & -0.829& -0.200 & -0.194 & 0.428 & 0.428 \\
			$B_{2,2}$  & 2.866  & 2.732 & -0.153 & -0.076 & 1.095 & 1.071 \\
			\hline
	 $g_z^{\rm gs}$    & 19.51  & 19.52 & 20.00  & 20.00  & 19.97 & 19.97\\
	 $g_x^{\rm mix}$   & 0.45   & 0.41  & 1.21   & 1.23   & 5.03  & 5.09 \\
	 $g_y^{\rm mix}$   & 0.82   & 0.74  & 8.84   & 9.36   & 7.34  & 7.19 \\
	 $g_z^{\rm mix}$   & 15.65  & 15.71 & 12.26  & 11.78  & 8.05  & 8.21 \\
			\hline
		$N_{\rm orb}$  & 259    & 722   & 262    & 897    & 261   & 785   \\
		    \hline
$T_{\rm CPU}^{\rm ROHF}$& \twocol{1.5}  & \twocol{5.2}    & \twocol{5.1} \\
$T_{\rm CPU}^{\rm CAS}$ & 0.15  & 0.8   & 0.23   & 1.7    & 0.20   &  1.38 \\
			\bottomrule
		\end{tabular*}
	\end{threeparttable}
\end{table}

\begin{figure*}[ht]
	\centering
	\subfigure[]{
		\label{fig:cmp-dBkqdr}
		\includegraphics[height = 5cm]{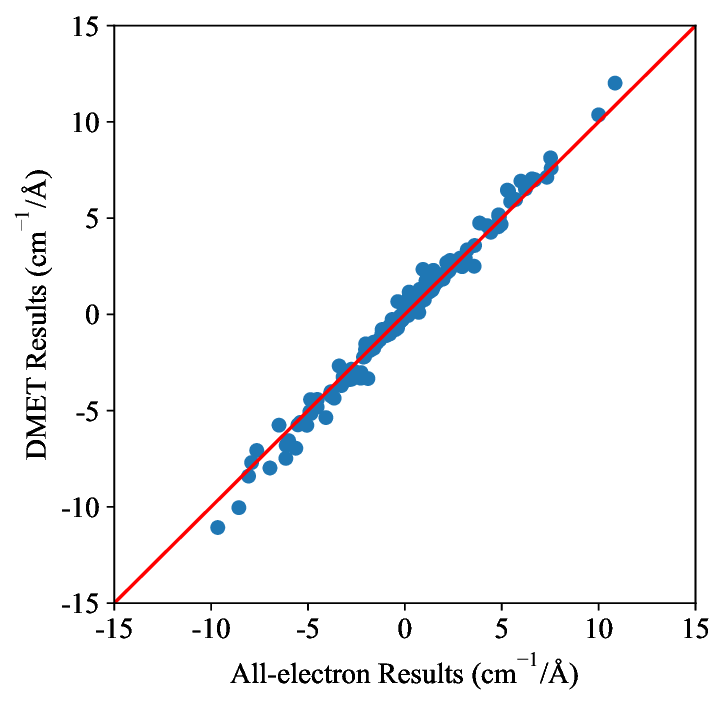}}
	\subfigure[]{
		\label{fig:cmp-rt}
		\includegraphics[height = 5cm]{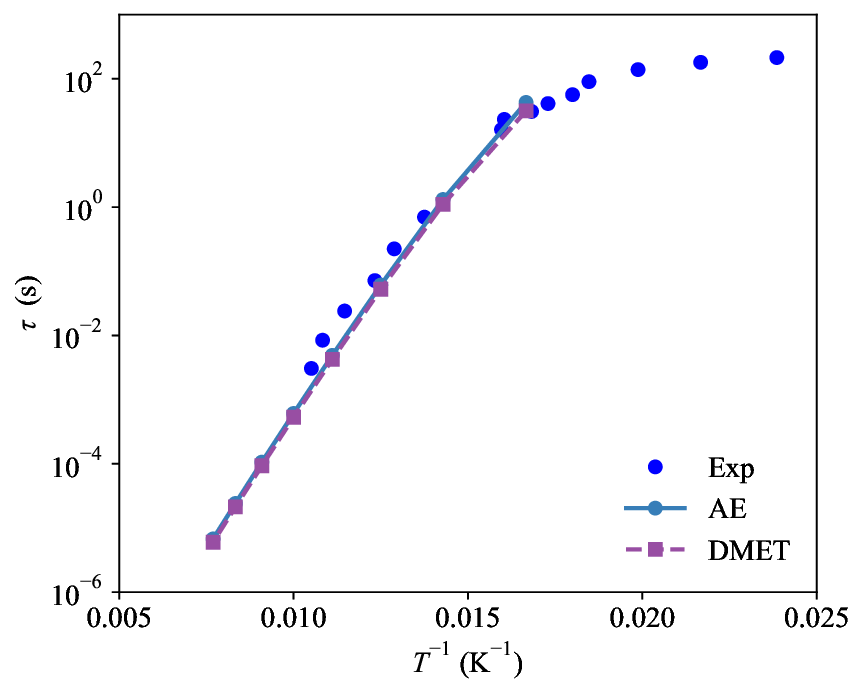}}
	\caption{(a) Comparison of all $\partial B_{kq}/\partial X_{i\alpha}$ values calculated by DMET and all-electron CASSCF-SO. (b) The relaxation times obtained from using SPC parameters computed via all-electron and DMET-based CASSCF-SO.}
	\label{Fig:diffrt}
\end{figure*}

\subsection{Validation of DMET+CASSCF-SO for static magnetic parameters}

In our previous study \cite{AiY2025}, we have demonstrated in three typical Ln-SIMs that DMET with the lanthanide atom treated as the impurity can well reproduce the all-electron description of crystal field splittings of the ground state multiplet of lanthanide ions at the CASSCF-SO level. To further validate the efficacy of DMET for Ln-SIMs, we first compare the results of static magnetic parameters $B_{kq}$ and $g$ tensors at equilibrium structures of three Ln-SIMs calculated by all-electron and DMET-based CASSCF-SO approach in Table \ref{tab:Bkq}. For $B_{kq}$, the results with $k=2$ are presented, since they play dominant roles in determining magnetic properties of Ln-SIMs. For effective $g$ tensors, we consider the results for the ground state and the lowest strongly mixed Kramers doublets (KDs), with the former determining the direction of the easy axis and the latter indicating magnetic states that contribute strongly to magnetic moment reversal. For all Ln-SIMs considered in this work, the ground state KD is strongly axial, exhibiting a large $g_z$ component and nearly vanishing $g_x$ and $g_y$, and therefore only the results for $g_z$ are presented. The results for the energies and the effective $g$ tensors of all KDs of these compounds are shown in \SI{Table S1-S3 in the Supplementary Material}. The DMET-derived energies are slightly higher than those from all-electron calculations, yet the relative deviations remain within 1-3\%, in agreement with earlier reports \cite{AiY2025}. For all three Ln-SIMs, $B_{kq}$ from DMET+CASSCF-SO agree very well with those from all-electron treatment, and the absolute errors are consistently below 0.2 cm$^{-1}$. Regarding the effective $g$ tensors, the two treatments yield nearly identical $g_z$ values for the ground state KDs. For the $g$ tensors of the lowest strongly mixed KDs, the errors are somewhat larger, but remain below 0.6, indicating excellent agreement between the $g$ tensors obtained from the two approaches. The dominant components of states (shown in \SI{Table S1-S3 in the Supplementary Material}) further indicate that the degrees of mixing in all KDs computed by the two methods are essentially identical. Taken together, these results demonstrate the reliability of the DMET-based CASSCF-SO approach for the prediction of magnetic parameters of a given structure.

\subsection{Validation of DMET+CASSCF-SO for SPC parameters}

We further check the accuracy of the DMET treatment of SPC parameters, which are calculated from the differences of $B_{kq}$'s of a series of off-equilibrium structures, and are therefore expected to be more sensitive to the numerical accuracy of the computational scheme. Taking \DyCp~as an example, we compare SPC parameters calculated with DMET+CASSCF-SO against those from all-electron calculations, and check the effects on the resultant spin relaxation time $\tau$. As illustrated in Fig. \ref{fig:cmp-dBkqdr}, the DMET SPC results exhibit very good agreement with the all-electron ones, with a mean absolute error of only about 0.011 cm$^{-1}$/\AA. This discrepancy has a negligible impact on the computed relaxation time as a function of temperature, as shown in Figure \ref{fig:cmp-rt}. Fitting the calculated $\tau$ vs $T$ in terms of
\begin{equation} \label{eq:tau-vs-T}
 \tau^{-1} = \tau_0^{-1} \exp{(-\Ueff/{k_{\rm B}T})},
\end{equation}
the values of $\Ueff$ from DMET and all-electron treatment differ by only 12 cm$^{-1}$, and those of $\lg(\tau_0/s)$ differ by 0.01, with the relative errors of only 1\% and 0.1\%, respectively. It should be noted that the good agreement between DMET and all-electron results in SPC and the resultant relaxation time observed above are obtained in spite of the fact that they were obtained by using different programs (PySCF vs ORCA) with different relativistic treatment and basis sets. Due to the high computational cost, we did not perform all-electron CASSCF calculations for SPC parameters within the PySCF framework. One can therefore expect that if all other settings of these calculations are identical, the resulting SPC parameters would be even in better agreement.

Before moving on, we comment on the computational savings gained by using DMET compared to all-electron treatment. For the systems considered in this work, the first step of the calculation, ROHF, is shared by both DMET and all-electron CASSCF-SO, which is used to build the embedded impurity space in the former, and prepare initial orbitals for the latter, and typically takes about a few hours (i.e. $T_{\rm CPU}^{\rm ROHF}$ in Table \ref{tab:Bkq}), which is more time-consuming than the CASSCF part for the current choice of the active space, i.e. CAS(9e,7o). Using DMET can greatly reduce the computational cost of CASSCF ($T_{\rm CPU}^{\rm CAS}$ in Table \ref{tab:Bkq}) since the orbital variation is conducted in a much smaller space compared to the all-electron calculation as indicated by the number of orbitals in the embedded impurity space (i.e. $N_{\rm orb}$ in Table \ref{tab:Bkq}). Apparently DMET leads to little CPU time saving in a single CASSCF-SO calculation. However, when calculating SPC parameters that requires conducting CASSCF-SO calculations for hundreds of structures around the equilibrium structure, the converged ROHF solution for one structure can be used as the initial guess for ROHF calculations of nearby structures, which can dramatically reduce the computational cost of subsequent ROHF calculations. Consequently, DMET can substantially lower the computational cost of SPC parameter evaluations.

\begin{figure}[ht]
	\includegraphics[width = 0.5\textwidth ]{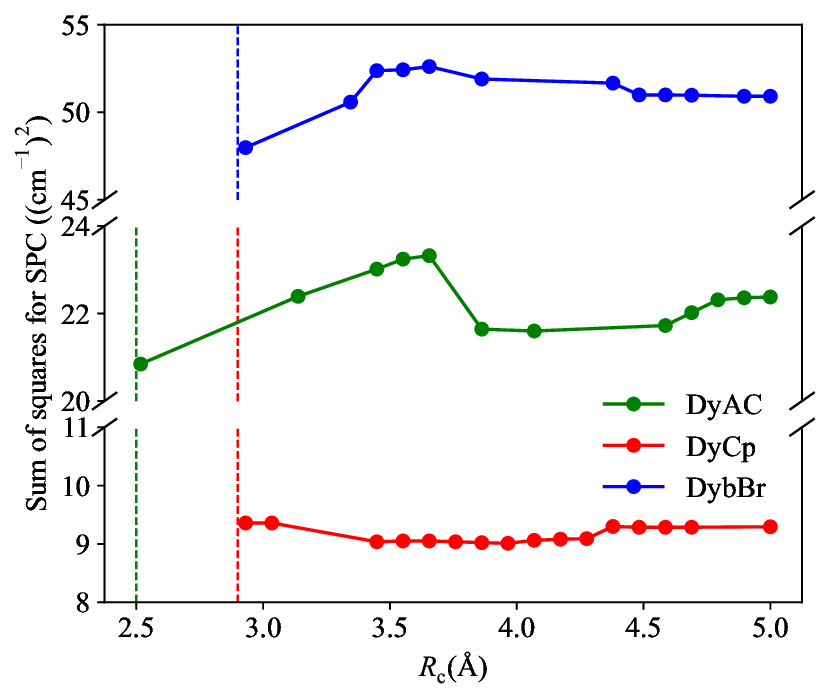}
	\caption{The sum of squares for SPC $\sum_{\nu} \sum_{kq}|B_{kq,\nu}(R_\mathrm{c})|^2$ for \DyAC~(green), \DyCp~(red) and \DybBr~(blue) as a function of $R_{\rm c}$. The vertical dashed lines mark the radius of the first coordination sphere for each system (green: \DyAC, 2.5 \AA; red: \DyCp, 2.9 \AA; blue: \DybBr, 2.9 \AA).
	}
	\label{Fig:dSPCdr}
\end{figure}

\subsection{Validity of spatial truncation}\label{Sec:space}

We further validate the effects of the spatial truncation on the accuracy of SPC parameters and spin relaxation simulation. Figure \ref{Fig:dSPCdr} presents the sum of squared truncated SPC parameters, i.e. $\sum_{\nu}\sum_{kq}|B_{kq,\nu}(R_\mathrm{c})|^2$, as a function of the spatial cutoff for \DyAC, \DyCp~and \DybBr. It can be observed that for all these three systems, the SPC parameters are predominantly contributed by atoms within the first coordination sphere. For \DyCp, once $R_\mathrm{c}$ exceeds its first coordination sphere (2.9 \AA), the contribution becomes negligible. For \DyAC, although atoms beyond the first coordination sphere (2.5 \AA) exert some influence, their contributions largely cancel each other; consequently, the SPC magnitude computed at $R_\mathrm{c} = 2.5$ \AA\ is already comparable to that obtained when all atoms are included. In contrast, for \DybBr, atoms located beyond the first coordination sphere (2.9 \AA) make non-negligible contributions to the SPC parameters.

\begin{figure}[ht]
	\centering
	\subfigure[]{
		\label{Fig:Ueffdr}
		\includegraphics[width=0.5\textwidth]{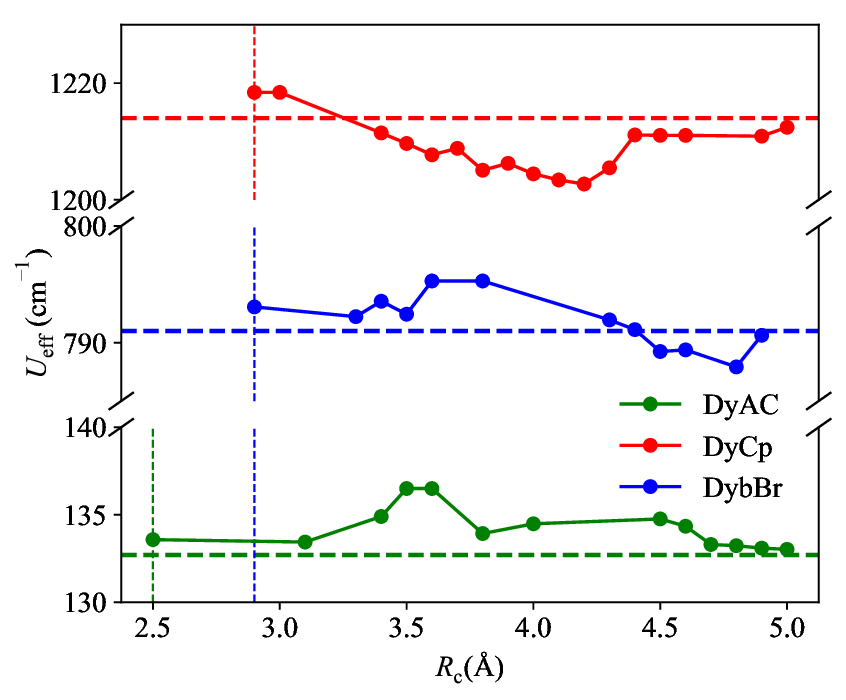}}
	\subfigure[]{
		\label{Fig:taudr}
		\includegraphics[width=0.5\textwidth]{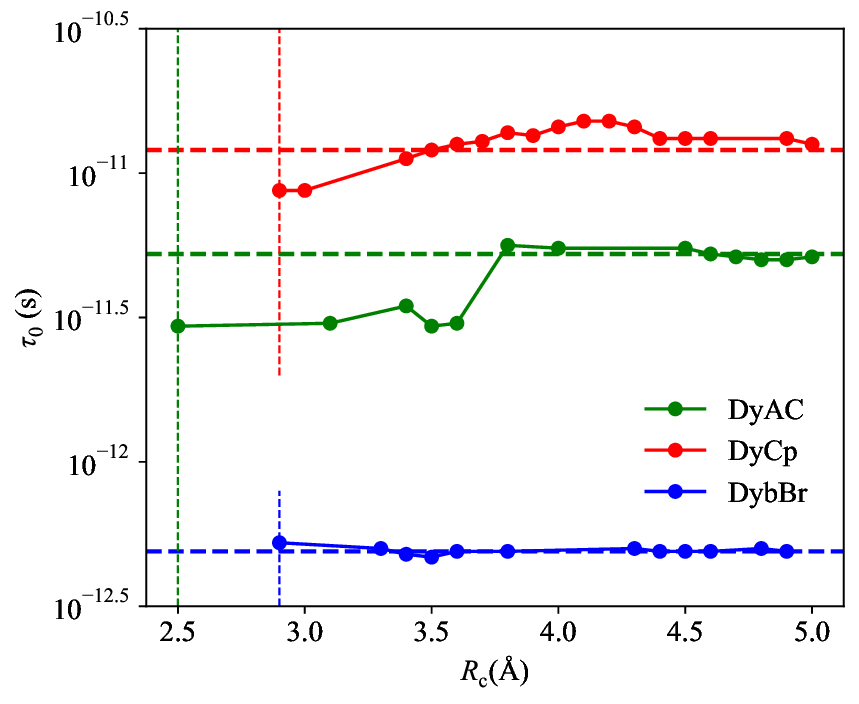}}
	\caption{
	    Calculated $U_{\mathrm{eff}}$ (upper) and $\tau_0$ (down) obtained with the truncated SPC parameters $B_{kq,\nu}(R_{\mathrm{c}})$ defined by Eq.(\ref{Con:SPCCdr}) as a function of the spatial cutoff $R_{\rm c}$. The dashed horizontal lines represent $U_{\mathrm{eff}}$ and $\tau_0$ obtained using the SPC parameters without spatial truncation, while the vertical dashed lines mark the corresponding ranges of the first coordination sphere for each system. Green: \DyAC, 2.5 \AA; red: \DyCp, 2.9 \AA; blue: \DybBr, 2.9 \AA.}
	\label{Fig:Uefftaudr}
\end{figure}

To investigate the effects of the spatial truncation on spin relaxation process, we calculate the relaxation time as a function of temperature using spatially truncated SPC parameters $B_{kq,\nu}(R_{\mathrm{c}})$ with different spatial cutoff $R_{\rm c}$, and by fitting $\tau$ vs $T$ using Eq.\ref{eq:tau-vs-T}, we obtain $U_{\rm eff}$ and $\tau_0$ as a function of $R_{\rm c}$, as shown in Fig. \ref{Fig:Uefftaudr}. The results show that, as $R_\mathrm{c}$ increases, both $U_{\mathrm{eff}}$ and $\tau_0$ converge to the reference values obtained from the calculation with SPC parameters derived from all atoms. When the truncation is limited to the first coordination sphere, the absolute relative error (ARE) in the calculated $U_{\mathrm{eff}}$ and $\lg{(\tau_0 /\mathrm{s})}$ for all systems are within 0.2-2.1\% of the converged values. At this truncation level, the ratios of the number of atoms included in the truncated region to the total number of atoms are 9/67 = 13.4\% for \DyAC, 11/93 = 11.8\% for \DyCp, and 8/64 = 12.5\% for \DybBr, indicating that the spatial truncation reduces the computational cost by approximately one order of magnitude. 

\begin{table}[ht]
	\centering
	\caption{The effective energy barrier $U_{\mathrm{eff}}$ and relaxation time-scale $\lg{(\tau_0 /\mathrm{s})}$ calculated using the spin-phonon coupling (SPC) parameters obtained via different methods for three systems are presented. In the table, `Expt' denotes the experimental results taken from Refs.\citenum{JiangSD2010}, \citenum{Goodwin2017} and \citenum{Liu2016} for \DyAC, \DyCp~and \DybBr, respectively, `AE' and `DMET' refers to the results using the all-electron and DMET, respectively, `DMET(1st)' represents DMET results obtained from using spatially truncated SPC parameters with atoms selected from the first coordination sphere, and `DMET(se)' denotes using spatially truncated SPC parameters with atoms selected based on their non-negligible contribution to the SPC, as discussed in Sec.\ref{Sec:space}. The numbers in the parenthesis show the percentage absolute relative error (ARE) of the corresponding treatment with respect to `DMET'.}
	\begin{threeparttable}
	\begin{tabular*}{0.48\textwidth}{@{\extracolsep{\fill}}ccll}
			\toprule
			System &  Method       &$U_{\mathrm{eff}}$ (cm$^{-1}$)&$\lg(\tau_0 /\mathrm{s})$\\
			\midrule
			\multirow{4}{*}{\DyAC} & Expt           & 129.9       & -10.66   \\
			                       & DMET          & 132.7        & -11.28   \\
			                       & DMET(1st)     & 131.3 (1.1)  & -11.52 (2.1) \\
			                       & DMET(se)      & 132.1 (0.45) & -11.26 (0.18)\\
			\midrule
			\multirow{4}{*}{\DyCp} & Expt           & 1223        &-10.42    \\
			                       & AE            & 1226         & -10.93  \\
			                       & DMET          & 1214         & -10.92   \\
			                       & DMET(1st)     & 1219(0.41)   & -11.06(1.28) \\
			\midrule
			\multirow{3}{*}{\DybBr}& Expt          & 712.4        & -11.38 \\
			                       & DMET          & 791.0        & -12.31 \\
			                       & DMET(1st)     & 793.0(0.25)  & -12.28(0.24) \\
			\bottomrule
		\end{tabular*}
	\end{threeparttable}
	\label{tab:Ueff-tau0}
\end{table}

It is worth noting that the ARE in $\lg{(\tau_0 /\mathrm{s})}$ for \DyAC~is slightly larger, reaching 2.1\%. This finding can naturally connect with the atom-resolved analysis of Briganti et al.\cite{Briganti2021} for \DyAC, which revealed that molecular vibrations beyond the first coordination shell contribute significantly to SPC through electrostatic polarization of the ligand backbone. We performed a similar analysis in \SI{Fig. S2 in Supplementary Material} and found that the sp$^2$-hybridized carbon atoms of the acetylacetonate ligand indeed make a discernible contribution to the SPC. The distance between the central Dy ion and these carbon atoms is 3.7-3.8~\AA. Consequently, when the truncation radius is extended to 3.8~\AA, the calculated results become almost identical to the converged values, as further supported by the data in TABLE \ref{tab:Ueff-tau0}. We therefore performed particularly tailored spatial truncation for \DyAC~in which the truncated region included the first coordination sphere atoms together with the sp$^2$ carbons of the acetylacetonate ligand. The resulting $U_{\mathrm{eff}}$ and $\lg{\tau_0}$ values are in excellent agreement with the all-atom results. Nevertheless, although atoms beyond the first coordination sphere do contribute to the SPC parameters, once these contributions are propagated to the $U_{\mathrm{eff}}$ and $\lg{\tau_0}$, their net effect is small. As shown by the data in TABLE \ref{tab:Ueff-tau0}, for all the three Ln-SIMs considered here, the values of $U_{\rm eff}$ and $\lg{\tau_0}$ obtained with spatially truncated SPC parameters that consider only the first coordination sphere differ from those of the full atom treatment show an absolute relative error of less than 2\%, which falls well within an acceptable tolerance. Therefore, restricting $R_\mathrm{c}$ to the first coordination sphere when calculating SPC parameters offers an optimal balance between computational efficiency and predictive reliability. In TABLE \ref{tab:Ueff-tau0} we also collect experimental values of $U_{\rm eff}$ and $\lg{\tau_0}$ for the three Ln-SIMs. The agreement between theory and experiment is quite good, consistent with the findings in previous studies \cite{Lunghi2022NRC, Chilton2022}.

\section{Conclusions}\label{sec:conclusion}

In summary, we have benchmarked the performance of the DMET+CASSCF-SO approach for computing static magnetic parameters and spin‑phonon coupling parameters of lanthanide single‑ion magnets. For the investigated Dy$^{3+}$-based complexes, DMET with the lanthanide ion treated as the impurity reproduces all‑electron CASSCF-SO results for effective crystal‑field parameters and $g$‑tensors with small deviations. When combined with finite‑difference evaluation of displaced geometries, the DMET scheme yields spin‑phonon coupling parameters that closely match all‑electron reference values for the DyCp complex, leading to only minor errors in the derived effective barrier \(U_\mathrm{eff}\) and relaxation timescale \(\lg(\tau_0/s)\). We further introduced a spatial‑truncation scheme for spin‑phonon‑coupling parameters, where only Cartesian derivatives from atoms within a given cutoff radius are retained, while full phonon eigenvectors and vibrational frequencies are preserved. For the three Dy-SIM systems, restricting contributions to atoms within the first coordination sphere yields relative errors below 2.1 \% for \(U_\mathrm{eff}\) and \(\lg(\tau_0/s)\), while reducing the number of relevant atoms by roughly one order of magnitude. Even though atoms outside the first coordination shell may can contribute to spin‑phonon coupling, their net influence on the final relaxation observables remains modest for these studied systems. For cases where outer‑shell atoms give non‑negligible contributions, extending the cutoff radius or manually including selected ligand atoms can further improve numerical convergence. This work demonstrates that combining DMET quantum embedding with spatial truncation offers a practical route to lower the substantial computational overhead of ab-initio spin‑phonon relaxation simulations for lanthanide single-ion magnets. It can facilitate future high-throughput computational screening and rational design of high-performance SIM materials. Further developments, including testing on broader families of lanthanide and transition-metal single‑ion magnets, incorporating high-order vibronic coupling, and combining with machine‑learning workflows, will further expand the scope and predictive power of this simulation framework.

\begin{acknowledgements}
This work was supported by the National Natural Science Foundation of China (grant numbers:12234001, 22573003 and 22131003) and Guizhou Provincial Science and Technology Innovation Leading Talent Workstation (KXJZ2025049). The High-Performance Computing Platform of Peking University is acknowledged for providing access to computational resources.
\end{acknowledgements}

\begin{flushleft} \textbf{SUPPLEMENTARY MATERIAL}\end{flushleft}
See the Supplementary Material for the energy levels, magnetic anisotropy parameters, dominant components of states obtained from DMET and all-electron (AE) CASSCF-SO calculation and molecular geometry of the complexes considered in this work.

\begin{flushleft} \textbf{DATA AVAILABILITY}\end{flushleft}
The data that support the findings of this study are available from the corresponding author upon request.

\bibliography{refs-LnSIMs}

\end{document}